\documentclass{article}
\usepackage{amscd,amsmath,amssymb}
\def\DH{\rm I\kern-1.5pt\rm H\kern-1.5pt\rm I}

\def\DR{\rm I\kern-1.45pt\rm R}
\def\DC{\kern2pt {\hbox{\sqi I}}\kern-4.2pt\rm C}

\newcommand{\ba}{\begin{array}}
\newcommand{\ea}{\end{array}}
\newcommand{\be}{\begin{equation}}
\newcommand{\ee}{\end{equation}}
\newcommand{\bea}{\begin{eqnarray}}
\newcommand{\eea}{\end{eqnarray}}
\newcommand{\bi}{\begin{itemize}}
\newcommand{\ei}{\end{itemize}}

\newcommand{\btau}{\mbox{\boldmath $\tau$}}

\begin{document}
\thispagestyle{empty}
\begin{center}
{\bf \Large Hopf maps and Wigner's little  groups}\\
\vspace{0.5 cm} {\large Ruben Mkrtchyan$^1$,  Armen Nersessian$^2$ and
Vahagn Yeghikyan$^3$}\\[3mm]

$\;^1${\sl Yerevan Physics Institute, 2 Alikhanian Brothers St., Yerevan, 0036, Armenia}\\
 $\;^2${\sl Yerevan State University, 1 Alex Manoogian St., Yerevan,
0025, Armenia}
\end{center}

\begin{abstract}
We present the explicit formulae relating  Hopf maps with Wigner's little  groups.
 They, particularly, explain simple action of group on a fiber for the first and second Hopf fibrations,
 and present most simplified form for the third one.
 Corresponding invariant Lagrangians are presented,
 and their possible reductions are discussed.

\end{abstract}
\section{Introduction}
 Hopf maps  play distinguished role in modern theoretical physics:
numerous constructions and models are related with them. Hopf maps
are of the special importance in the supersymmetry
\cite{Baez,Baez1}, monopoles \cite{atiah}, and more generally, in
supergravity/string theories. Hopf maps are useful in the study of
the problems of classical and quantum (supersymmetric) mechanics as
well. They are useful for the construction of the mechanical systems
(including supersymmetric ones) with monopoles by the reduction
method, including supersymmetric mechanics. Particularly, the
reduction related with the zero Hopf map yields the system with
anyons, with the first Hopf map - to the systems with Dirac
monopole, with the second Hopf map - with SU(2) Yang monopoles (for
the review see, e.g. \cite{lnp} and refs therein). The detailed
scheme of such a Lagrangian reduction related with the first/second
Hopf maps, formulated in complex/quaternionic number's language has
been developed in \cite{gknty}. The reduction scheme based on the
third Hopf map, and respectively, on octonionics, is not developed
yet, to our knowledge.

The Hopf maps are closely related with supersymmetric theories in $(3+1)-$, $(5+1)-$ and $(9+1)$ dimensions.
 In this respect let us mention the paper written decade ago by  Bandos, Lukierski and Sorokin \cite{bandos}, where
massless superparticle model with tensorial central charges
has been proposed and quantized. The authors of that paper noted the relation of their construction with Hopf maps,
 and  also actually used the action of (subgroup of the) little  group  of momenta on corresponding spinor.
  Couple of years later one of the authors (R.M.) considered and explicitly constructed irreducible representations of
  tensorial Poincar{\'e} groups \cite{mkrtchyan,mkrtchyan1} by Wigner's little group method \cite{wigner}.
  Tensorial Poincar{\'e} groups are  usual Poincar{\'e}, supplemented with additional tensorial central charges,
  corresponding to different supersymmetry algebras. It appears that exactly as quantization  of particle in Minkowski space leads
  to irreps of Poincar{\'e} group, quantization of particle in tensorial spaces leads to the  irreps of tensorial Poincar{\'e} groups.
  From the point of view of irreps of tensorial Poincar{\'e}, Hopf maps appear in the problem of decomposition of that irreps w.r.t.
  the usual Poincar{\'e} subgroup of tensorial Poincar{\'e}. Namely, when considering tensorial Poincar{\'e} in dimensions
   $(3+1)-$, $(5+1)-$ and $(9+1)-$, corresponding to the minimal ${\cal N}=1$ supersymmetry algebras, taking a ``preon''
   representation (i.e. those, corresponding to maximal number of surviving BPS supersymmetry), and decomposing
   that representations w.r.t. the usual Poincar{\'e} (i.e. considering particle content of preon),
   we get the Hopf fibrations of (unit norm) spinor spaces. In this way one conclude that the particle
   content of preon representation appears to be a tower of particles with all  integer spins with multiplicity one.
Note that we use here a notion of ``preon'' in a sense, introduced in \cite{4bandos}, which denote a special configuration of tensorial central charges of supersymmetric theories, such that matrix of tensorial charges $Z_{\alpha\beta}$ has rank one, i.e. $Z_{\alpha\beta}\approx \lambda_\alpha \lambda_\beta$. As shown in \cite{4bandos}, all other configuration of branes can be considered as some combinations of different number of preons, which explains the name of that object.

  However, this tight connections between Hopf maps and the little  groups of tensorial Poincar{\'e} had not
   been fully presented and studied up to now,
  as well as its differential-geometric formulation, while the latter is common language in the study of
  (super)particle mechanics and in the field theory. Moreover, as we noticed above, the (octonionic)
  reduction related with the
  third Hopf map, doesn't exist yet, and it is   unclear, which sort of ``monopole'' should  appear
  in the resulting system after such a reduction.

  In the present paper we  present the  explicit differential geometrical description of the action of the
 $d=(3+1), (5+1), (9+1)$-dimensional little groups, as well as establish their direct  correspondence with  Hopf  maps.
 We also write down the metrics and one-forms, which are invariant under action of the little  groups.
Beside the above mentioned ``practical" interest, our constructions are of the academic importance in the part,
related with third Hopf map.

It is worth to mention hear the difference with the (closely related) twistorial approach to the same objects - Hopf maps, spinorial manifolds (twistors), etc (see a review on connection of twistors in different dimensions, division algebras and Hopf maps in \cite{ced1}, and connected works \cite{ced2}). As is clear from above, and will be exploited below, we are concentraited, first, on a connection between little groups and  Hopf maps, actually obtaining the last object and its properties from the group theory approach, and, second, on a finite-dimensional mechanical systems and their reductions (as \cite{lnp,gknty}), based on corresponding manifolds.

\section{General consideration}
The description of Hopf maps can be given in terms of little groups (or stability groups) of Lorentz spinors.
 The generators of Lorentz group are given by $d\times d$-dimensional antisymmetric matrices $\omega_{\mu\nu}$ and
 the transformation of vector $V_\mu$ is given by the following expression:
\be
\delta V_\mu=\omega_{\mu\nu}V^\nu,\quad \mu,\nu=1,\ldots,d.\label{transL}
\ee
The transformation of corresponding spinor $\lambda$ is given via
$2^{d/2}$-dimensional matrices $\gamma^\mu$: \be
\delta\lambda=\omega_{\mu\nu}S^{\mu\nu}\lambda,\quad
S^{\mu\nu}=\frac{1}{2}[\gamma^\mu,\gamma^\nu],\quad
\{\gamma^\mu,\gamma^\nu\}=2g^{\mu\nu}, \label{tl}\ee where $g^{\mu\nu}$ is the
Minkowski's metrics.
The operators $S^{\mu\nu}$ form  $so(1,d-1)$ Lorentz algebra.

For some very special dimensionalities $d$ the  space of spinors (without zero) of $d$-dimensional Lorentz group $SO(1,d-1)$  can be represented as
factor space of this group by stability subgroup of one given spinor:
\be
\{\lambda\}=SO(1,d-1)/{\cal G}, \quad g \lambda_0=\lambda_0,\quad  g\in {\cal G},\lambda_0 \in\{\lambda\}
\ee
Calculations of these subgroups and dimensionalities can be found in \cite{mkrtchyan}, here we are interested only in the results of that paper for fixed dimensions where Hopf maps exist, so from now on
$ d=3+1, 5+1, 9+1$. The corresponding stability groups are:
\bea
&{\cal G}_{1,3}&= T^2\\
&{\cal G}_{1,5}&=SO(3) \ltimes T^4\\
&{\cal G}_{1,9}&=SO(7) \ltimes T^8,
\eea
where $T^k$ is $k$-dimensional translational subgroup of corresponding Lorentz group.
Here $\ltimes$ denotes semidirect product.

Now for every spinor $\lambda$ one can define $d$ quantities $p^\mu$ (components of vector) which will transform as in \ref{transL},
\be
p^\mu=\bar\lambda\gamma^\mu\lambda,\quad \bar\lambda =\lambda^{*} A,
\label{pl}\ee
where $*$ means complex conjugation and  $A=\gamma^0$. From this expression it follows, that  the vector $p^\mu$ is light-like one,
\be
p_\mu p^\mu=0.
\ee
On the other hand, one can consider all Lorentz  transformations which leave invariant the quantities $p^\mu$,
i.e. the matrixes $\omega_{\mu\nu}$, which satisfy an equation
\be
\omega_{\mu\nu}p^\nu=0.\label{pinv}
\ee
Resolving this equation, we get
\be
\left\{\begin{array}{ccc}
                            \omega_{0i}p^i=0\\
\omega_{i0}p^0+\omega_{ij}p^j=0
                            \end{array}
\right.\Rightarrow
\omega_{0i}=-\omega_{i0}=\frac{\omega_{ij}p^j}{p_0} \label{po}\ee
 It is obvious that above considered little groups of spinors are subgroups of these transformations.
 Indeed, any transformation, which does not change spinors $\lambda$ will also leave invariant vector-quantities $p^\mu$.
  For the considered dimensions this transformations are:
\bea
& d=3+1: & SO(2) \ltimes T^2\\
&d=5+1 & SO(4)\ltimes T^4\\
& d=9+1 & SO(8)\ltimes T^8
\eea

Which transformations change the spinor $\lambda$ but leave
invariant the quantities $p_\mu$?  It is clear, that they  form the bundle of fibration of spinor
space over the space of quantities $p^\mu$. For the considered cases we have
\bea
SO(2) \otimes T^2/T^2=SO(2)\equiv S^1\nonumber\\
SO(4)\otimes T^4/SO(3) \otimes T^4=SO(3)\equiv S^3\label{factor}\\
 SO(8)\otimes T^8/SO(7) \otimes T^8=S^7\nonumber
\eea
One can  prove (at least, for the  special representation of $\gamma^\mu$ matrices)
that fixation  $2^{d/2}-1$-dimensional sphere in the space of spinors leads to fixation of  $(d-2)$-dimensional sphere
in the space of $p^\mu$:
\be
(\lambda^*)^T\lambda=p_i p_i=1,\quad i=1,\ldots, d-1\label{ninv}.
\ee
Let us notice,  that this identity is not Lorentz-invariant one,  and depends on the realization of matrices $\gamma^\mu$.
 On the other hand, the existence of this identity in one realization leads to the existence of similar identity
 in any other one, which can be obtained by unitary transformation. Because of non-invariant form of (\ref{ninv}),
 the points of spheres in different realizations are not coincide.

So, taking in mind the bundles (\ref{factor}) we get  factorizations of spheres over spheres,
\be
S^3/S^1=S^2,\qquad
S^7/S^3=S^4\qquad
S^{15}/S^7=S^8,
\label{hm0}\ee
i.e. the first, second and third Hopf maps.
We presented  quite simple procedure for the describing Hopf maps.
However, the considered transformations are too general and not so easy to use.
Indeed,  the matrices $\omega_{\mu\nu}$  obey  the condition (\ref{pinv}),
 which  decreases the number of $d(d-1)/2$ parameters by $d-1$ only.
 This matrix depends on the components $p_\mu$ and, therefore, on spinor $\lambda$. Hence,   transformations (\ref{transL})
  are non-linear ones. On the other hand, any linear transformation  $\omega \mapsto f(\omega,\lambda)$,  where $f$ is an arbitrary
 function, does not change  the condition (\ref{pinv}).

Whether it is possible to choose the form of $\omega_{\mu\nu}$ simplifying  the above  transformation (\ref{tl}) to the linear one?
For answering on this question, in the next sections we shall consider the transformation (\ref{tl}) separately for the
$d=3+1, 5+1, 9+1$  dimensions.
However, before going to do that, we will consider, in the next section, the relation of the Hopf map with division algebras.

\section{Hopf maps}
The Hopf maps $ {  S}^{2n-1}/{  S}^{n-1}={  S}^{n}$,
$n=1,2,4,8$ are closely related with the division algebras.
Precisely, zero Hopf map $n=1$ reflects the existence of division algebra of real  numbers, first Hopf map $n=2$  - complex numbers,
 second Hopf map $n=4$ - quaternions,
third Hopf map $n=8$ -octonions.

Let us describe the  Hopf maps in  explicit terms, demonstrating their relation with the division algebras
For this purpose, we consider the functions
\be {\bf p}=2{\bf \bar u}_1{\bf u}_2 ,\quad p_{n+1}={\bf \bar
u}_1{\bf u}_1-{\bf \bar u}_2{\bf u}_2, \label{hm}\ee
where ${\bf
u}_1,{\bf u}_2$ are real numbers for the $n=1$ case (zero Hopf map),  complex numbers for the $n=2$ case (first Hopf map),
 quaternionic numbers for the $n=4$ case (second Hopf map) and octonionic numbers  for the $n=8$ case (third Hopf map). One
can consider them as coordinates of the $2n$-dimensional  space
$\DR^{2n}$ ( $n=1$ for   real ${\bf u}_{1,2}$; $n=2$  for complex ${\bf u}_{1,2}$; $n=4$ for
quaternionic ${\bf u}_{1,2}$; $n=8 $ for octonionic ${\bf u}_{1,2}$).
In all  cases  $p_{n+1}$ is a
real function, while ${\bf p}$ is, respectively, real function ($n=1$),  complex function
($n=2$),  a quaternionic one ($n=4$), and octonionic one ($n=8$),
\be {\bf p}\equiv p_{n}
+\sum_{k=1,\ldots, n-1}{\bf e}_k p_k,\qquad {\bf e}_i{\bf e}_j=-\delta_{ij}+ C_{ijk}
{\bf e}_k\ee
where the structure constants $C_{ijk}$ are totally antisymmetric by indices $(ijk)$,
so that ${\bf e}_k \equiv 0$ for $n=1$;
${\bf e}_k \equiv
{\bf i}$, ${\bf i}^2=-1$,  $c_{ijk}=0$ for $n=2$;   ${\bf e}_k\equiv ({\bf i},{\bf
j},{\bf k})$, $C_{ijk}=\varepsilon_{ijk}$ for $n=4$.
For $n=8$  the structure constants $c_{ijk}$ are defined by the relations
\be
C_{123}=C_{147}=C_{165}=C_{246}=C_{257}=C_{354}=C_{367}=1,
\ee
while all other non-vanishing components are determined by the total antisymmetry.
 Hence, $(p_{n+1}, {\bf p})$ parameterize the $(n+1)$-dimensional space $\DR^{n+1}$.

 One can check immediately,  that the following equation holds: \be p^2_0\equiv {\bf \bar
p}{\bf p}+p^2_{n+1}=( {\bf\bar u}_{1}{\bf u}_1+{\bf \bar u}_2{\bf
u}_2)^2. \label{hm1}\ee
 Thus, defining the $(2n-1)$-dimensional sphere with radius  $\sqrt{p_0}$, ${\bf\bar u}_\alpha{ \bf u}_\alpha =p_0$ in
$\DR^{2p}$,
we will get the $p$-dimensional sphere with radius
$p_0$ in $\DR^{p+1}$ .

The expressions (\ref{hm}) can be easily inverted by the use of equality (\ref{hm1}) \be
{\bf u}_\alpha={\bf g} r_\alpha ,\qquad {\rm where }
\quad r_1=\sqrt{\frac{p_0+p_{p+1}}{2}},\quad  r_2\equiv
r_+=\frac{{\bf p}}{\sqrt{2(p_0+p_{p+1})}},\quad
,\quad {\bar {\bf g}}{\bf g}=1 .
\label{inve}\ee
It follows from the last equation in (\ref{inve}) that ${\bf g}$
parameterizes the $(p-1)$-dimensional sphere of unit radius.\\

Using  above equations, it is easy to describe the first three Hopf maps. Indeed, for
$n=1,2,4$
the functions  ${\bf p}, p_{n+1}$ remain invariant under the
 transformations
\be {\bf u}_\alpha\to {\bf{G}} {\bf u}_\alpha ,\quad {\rm
where}\quad {\bar {\bf {G}}}{\bf{G}}=1 
\label{G}\ee Therefore,  ${\bf G}$ parameterizes the spheres $S^{n-1}$ of
unit radius. Taking into account the  isomorphism
 between these spheres and the groups for $n=1,2,4$ ($S^0=Z_2$, ${  S}^1=U(1)$, ${  S}^3=SU(2)$),
we get that (\ref{hm})  is invariant under $G-$group
transformations for $n=1,2,4$ (where $G=Z_2$ for $n=1$,  $G=U(1)$ for $n=2$, and $G=SU(2)$ for
$n=4$).

It is easy to construct the one-forms and metrics, which are invariant under ``global"  $G$-action (\ref{G}),
i.e. for  ${\bf G}=const$:
\be
\omega_1^0={\rm Re}\; f_\mu ({\bf p})({\bf \bar u}\sigma^\mu d{\bf u}) +
{\rm Re}\; \lambda_{\mu\nu}(p)({\bf \bar u}\sigma^\mu {\bf\bar u})({\bf u}\sigma^\mu d{\bf u}),
\label{1formGlob}\ee
\be
(ds)^2=f_{\mu}(p)(d{\bf\bar u}\sigma^\mu d{\bf u})+
f_{\mu\nu}(p)({\bf\bar u}\sigma^\mu d{\bf u})(d{\bf\bar u}\sigma^\mu {\bf u})+
{\rm Re}\;\lambda_{\mu\nu}(p)({\bf\bar u}\sigma^\mu{\bf\bar u})(d{\bf u}\sigma^\mu d{\bf u}),
\label{metricGlob}\ee
where $ f_{\mu\nu}(p)=\overline{f}_{\nu\mu}(p)$, and $f_\mu (p),\lambda_{\mu\nu}$ are arbitrary complex functions.

The Lagrangians constructed by the use of the above one-form and metric, will also be $G$-invariant one. Using Hopf maps,
 they can be easily reduced to the low dimensional systems both in the Hamiltonian and Lagrangian framework. It is worth to mention here that spinorial origin of quantities in Lagrangians doesn't lead to any spin-statistics problems both in this non-relativistic case (where one don't expect them due to non-relativistic framework), as well as in relativistic case, see e.g. \cite{bandos}, where bosonic spinors essentially play the role of source of constraints.

For  $n=1$ case ($G=Z_2$, zero Hopf map)  we simply reformulate
two-dimensional  system in terms  $Z_2$-invariant coordinates, i.e. perform
(locally) equivalence transformation. Nevertheless, the  reduction
by $Z_2$ group has the consequences if we consider global properties
of the system. It becomes visible on the quantum-mechanical level, and
yields the system interacting with  the anyons (see, e.g.,  \cite{ntt}).

For the $n=2$ case ($G=U(1)$, first Hopf map) the reduction results the initial four-dimensional system
in the three-dimensional one interacting with Dirac monopole. For the particular examples of such a (Hamiltonian)reductions,
 we refer to \cite{U1,U12}
 and references therein.

For the $n=4$ case ($G=SU(2)$, second  Hopf map) the reduction is more complicated, due to non-Abelian nature
 of $SU(2)$ group. It results the initial eight-dimensional system in the five-dimensional one, interacting with $SU(2)$ Yang monopole.
 respectively, the reduced system has  twelve-dimensional  phase space $T^*R^5\times S^2$.
 For the particular example  of such a Hamiltonian reduction
 we refer to \cite{casteill,casteill1} and references therein.
Let us also mention the recent paper \cite{gknty}, where the detail procedure of the Lagrangian reduction related with first and
second Hopf map has been developed  on the simplest example of the
$2n$-dimensional  Lagrangian
$
{\cal L}_{0}=g({\bf \overline u}\cdot {\bf
u}){\bf\dot{\overline u}}_\alpha {\bf\dot{u}}_\alpha $. Its extension to the Lagrangians associated with metric
 (\ref{metricGlob}) is straightforward.\\

For the octonionic case $n=8$ situation is more complicated.
Because of losing associativity the standard
transformation of ${\bf u}_\alpha$ that leaves invariant
coordinates ${\bf p}, p_9$ will not be just (\ref{G}).
Its modification can be easily obtained using
(\ref{inve}): \be {\bf u}_\alpha \mapsto ({\bf G}{\bf g})({\bf
\bar g}{\bf u}_\alpha)= \frac{({\bf G}{\bf u}_1)({\bf
\bar u}_1{\bf u}_\alpha)}{{\bar{\bf u}_1 {\bf u}_1}} . \label{octtrans} \ee
Also, for the $n=8$ case the fiber $S^7$ is not isomorphic  with any group.
So, it is unclear, how to construct the
octonionic analog of the one-form (\ref{1formGlob}) and metric (\ref{metricGlob}), which will be invariant under transformation
(\ref{octtrans}).
So, one can expect further troubles in the extensions
 of the constructions, related with the lower Hopf maps to the third one.
As a consequence, the extension of the above-described reduction procedures to the octonionic case is an open problem yet.

\section{$d=3+1$ case and first Hopf map}
Let us consider the action of the Wigner's little group of momenta on the corresponding spinors,  in the $d=3+1$ dimension.
For this purpose it is  convenient to choose a specific realization of  $\gamma$-matrices,
\be
\gamma^i=\left(
\begin{array}{cccc}
 0 & \sigma^i \\
 (\sigma^i)^T & 0
\end{array}
\right),\quad \gamma^0=\left(
\begin{array}{cccc}
 0 & {\bf 1}_2\\
 -{\bf 1}_2 & 0
\end{array}
\right),\quad A=\gamma^0,\quad i=1,2,3.
\ee

Upon this realization,
 for the
Majorana spinor $\lambda$ the expressions for momenta (\ref{pl}) read \be
p^\mu=\lambda^*\gamma^0\gamma^\mu\lambda,\label{pdef} \ee 
where
\be \lambda= \left(
\begin{array}{ccc}
 z_1\\
z_2\\
\bar z_2\\
-\bar z_1
\end{array}
\right) \ee
Hence, one can replace the expressions  (\ref{pdef}) with the following
ones: \be p^0=Z\bar Z,\quad p^i=Z\sigma^i  \bar Z,\quad
\lambda=(Z,\sigma^1\bar Z)\quad Z=(z_1,z_2),\quad \bar Z= Z^* \ee
In that case the solutions of the equations (\ref{po}) can be represented as follows
: \be
\omega_{0i}=\epsilon_{ijk}a^jp^k,\quad
\omega_{ij}=\epsilon_{ijk}a^kp^0,\label{lucum} \ee
while corresponding
transformation of spinor $\lambda$  (\ref{tl}),  will take a form  \be
\delta\lambda=\omega_{\mu\nu}\gamma^{\mu\nu}\lambda=2\omega_{0i}\gamma^{0i}+\omega_{ij}\gamma^{ij}.
\ee
Here \be
\gamma^{0i}=\frac{\left[\gamma^0,\gamma^i\right]}{2}=\left(\begin{array}{ccc}
                -\sigma^i & 0\\
0 & -\sigma^i
               \end{array}
\right), \quad
\gamma^{ij}=\frac{\left[\gamma^i,\gamma^j\right]}{2}=\left(\begin{array}{ccc}
                -\imath\epsilon^{ijk}\sigma_k & 0\\
0 & -\imath\epsilon^{ijk}\sigma_k
               \end{array}
\right)=\imath\epsilon^{ijk}\gamma_{0i}. \ee

Equivalently,
\be \delta
Z=2\omega_{0i}\sigma^iZ+\omega_{ij}\epsilon^{ijk}\sigma_kZ \ee
Substituting the solutions from (\ref{lucum}) in the second  summand, and transforming the first one by the use of
Fierz identity\be
i\epsilon^{ijk}(u\sigma_j \bar v)(r\sigma _k \bar s)=(u\bar
s)(r\sigma_k\bar v)-(r\bar v)(u\sigma_i\bar s), \ee
we get  \be \delta Z=\imath
a^i\; (Z\sigma_i\bar Z)Z \label{1f4}\ee
Since $a_i$ are arbitrary functions depending on $Z,\bar Z$, we can
represent (\ref{1f4}), upon their proper  redefinitions, as follows
 \be \delta Z^\alpha=\imath a(Z, \bar Z)Z^\alpha
\label{2f4}, \ee
with $a(Z,\bar Z)$ being arbitrary function.
The
respective finite transformation reads $Z\to {\rm e}^{\imath
\varphi(Z, \bar Z)} Z$. This is a phase shift, so we get that the
little group of momenta on corresponding four dimensional Lorenz spinor is $U(1)= S^1$,
parametrising the sphere of unit radius. The action of the little
group has a natural geometric interpretation in terms of first Hopf
map. Indeed, the relation
 ${\bar Z} Z=\sqrt{p_i p_i}$,, upon fixing of value of ${\bar Z} Z=p_0=const$,  yields
 the fibration corresponding to the first Hopf map $S^3/S^1=S^2$ (see previous
 Section). So, the spinors can be formulated in the coordinates associated with Hopf fibration
 (\ref{inve}), viz
 \be
 Z^\alpha=\left({\rm e}^{\imath\gamma}\sqrt{\frac{p_0+p_3}{2}},\quad
 {\rm e}^{\imath\gamma}\frac{(p_1+\imath p_2)}{\sqrt{2(p_0+p_3)}}\right),
 \ee
where the angle $\gamma\in [0,2\pi)$ parameterizes the  fiber $S^1$.
Thus, the transformation (\ref{2f4}) acts on the fiber coordinate $\gamma$ only: $\gamma\to \gamma+a(\gamma, p)$.

There is variety of ways to construct the Lagrangians which are
invariant under this, ``local" $U(1)$ transformations.

For example, one can construct the one-forms, which are invariant
under (\ref{2f4}): \be \omega^1= \imath f_{j}({\bar Z}\sigma^j d Z)
+\lambda_{ij}(p)(\bar Z\sigma^i \bar{Z})(Z\sigma^j d{Z})+\quad c.c.,
\label{ho2}\ee where $f_{i}(p)$ and $\lambda_{ij}(p)$ are real
symmetric tensors.The respective first-order Lagrangian will be
invariant under above transformation as well.

One can also construct the  metric which is invariant with respect to (\ref{2f4})transformation: \be
ds^2=ds^2_{0}+ds^2_{1}, \label{ds}\ee where \bea && ds^2_{0}
=\imath b_{ij}(p)\left[ p^i (d{\bar Z}\sigma^j d Z)-
(d{\bar Z}\sigma^i Z)({\bar Z}\sigma^j d Z)\right], \qquad b_{ij}(p)=b_{ji}(p)\label{h}\\
&&
ds^2_{1}= f_{lk ij}\left[
{\rm Re} \; (\bar Z\sigma^l \bar{Z})(\bar Z\sigma^k \bar{Z})(Z\sigma^i d{Z})(Z\sigma^j d{Z})
 -2
( Z\sigma^l {Z})(\bar Z\sigma^k \bar{Z})(Z\sigma^i d{Z})(\bar
Z\sigma^j \bar d{Z})\right], \label{hol}\eea  where
$f_{lkij}(p)=f_{kl ij}(p)$, $
f_{lk ij}(p)=f_{lk ji}(p)$, and is $p^i$ expressed via
$\bar Z$, $Z$  in accordance with (\ref{pdef}).
 It is clear, that by
the use of this metric we can construct reparametrization-invariant Lagrangian
${\cal L}=|ds_/dt|$ and sigma-model-type Lagrangian as well, ${\cal
L}=ds^2/dt^2$. One  can also construct the
higher-derivative Lagrangians, which are invariant under
(\ref{2f4}): these Lagrangians will be the functions of extrinsic
curvatures associated with the above constructed metrics. Surely, such
 generic Lagrangians are not invariant under action of the whole
 Lorentz group, but under its little group only.

It is easy to see, that both  particular metrics, (\ref{h}) and
(\ref{hol}) as well as their sum (\ref{ds}), are degenerated one:
  the vector field
 ${\bf V}=Z^\alpha \partial_\alpha +c.c.$ defines their null vector. Hence, the respective Lagrangians will be
degenerated: they will have constraint defined by this null
vector. In fact, these Lagrangians could be reformulated as a
non-singular ones  depending on momenta ${\bf p}$ only.

On the other hand, instead of generic local transformation (\ref{2f4}) one can consider particular one, when
$a\equiv p_ia_i(Z,\bar Z)$ is a constant number. For this purpose one can choose, e.g
\be
a_i(Z,\bar Z)=\frac{c_i}{p_i}\;({\rm with}\;{\rm  no}\;{\rm  summation}\;{\rm  over}\; i ),\qquad c_i={\rm const}.
\label{flat4}\ee
In that case the  action of the little group on the spinor will be multiplication on the constant phase factor.
The reduction of the four-dimensional systems under such $U(1)$
group action is widely known in classical and quantum mechanics, and yields the systems interacting with
 Dirac monopole (see, e.g. \cite{lnp,gknty}).

\section{$d=5+1$ case and second Hopf map}

In this Section we consider the action of little group of momenta on the corresponding spinors in $d=5+1$ dimensional space.
Since in $d=5+1$ case there is no Majorana spinor, we should use a Weyl
one, \be \lambda=(z_1,z_2,z_3,z_4,0,0,0,0).
 \ee
 Similar to the previous case,  we define \be
p^\mu=\bar\lambda\Gamma^\mu\lambda,\qquad {\rm where }\quad \bar
\lambda=\lambda^*\Gamma^0,\quad  \Gamma^0=\left(
\begin{array}{ccc}
 0 & \imath{\bf 1}_4\\
-\imath{\bf 1}_4 & 0
\end{array}
\right),\quad \Gamma^i=\left(
\begin{array}{ccc}
 0 &  \gamma^i\\
-\gamma^i & 0
\end{array}
\right),\quad i=1,\ldots,5.
\label{pdef4} \ee
Then we introduce  \be Z=(z_1,z_2,z_3,z_4),\quad \bar
Z=Z^*, \ee and rewrite the expression for momentum $p_\mu $ (\ref{pdef4}) in the following form: \be
p^0=Z\bar Z,\quad p^i=-\bar Z\gamma^i Z \label{pthz} \ee
 Now, for the
transformation for spinor $\lambda$ we have: \be
\delta\lambda=\omega_{\mu\nu}\Gamma^{\mu\nu}\lambda=2\frac{\omega_{ij}p^j}{p^0}\Gamma^{0i}\lambda+\omega_{ij}\Gamma^{ij},\label{dlambda}
\ee where \be \Gamma^{0i}=\frac{\left[\Gamma^0,\Gamma^i\right]}{2}=
\left(
\begin{array}{ccc}
-\gamma^{i} & 0\\
 0 & -\gamma^{i}
\end{array}
\right),\quad \Gamma^{ij}=\frac{\left[\Gamma^i,\Gamma^j\right]}{2}=
\left(
\begin{array}{ccc}
-\gamma^{ij} & 0\\
 0 & -\gamma^{ij}
\end{array}
\right), \ee where $\gamma^i$ are Euclidean gamma-matrices: \be
\left\{\gamma^i,\gamma^j\right\}=2\delta^{ij} \ee It is easy to see,
that one can replace the transformation formula for $\lambda$
(\ref{dlambda})  with the following one \be \delta
Z=-2\frac{\omega_{ij}(\bar Z\gamma^j
Z)}{p_0}\gamma^{i}Z+\omega_{ij}\gamma^{ij}Z \label{initd6}\ee
For further manipulations with this expression  we will use the Fierz identity
 \be Z_\alpha \bar
Z_\beta=\frac{1}{4}(Z\bar Z)\delta_{\alpha\beta}+\frac{1}{4}(\bar
Z\gamma^l Z)\gamma^l_{\alpha\beta}-\frac{1}{8}(\bar
Z\gamma^{lm}Z)\gamma^{lm}_{\alpha\beta} .\ee
Applying this identity  to (\ref{initd6}), we get, after  proper  simplification,
\be \omega_{ij}(\bar Z\gamma^j
Z)\gamma^{i}Z=-(\bar Z\gamma^{ij}Z)\omega_{ij}Z-(\bar
Z\gamma^{lj}Z)\omega_{ij}\gamma^{il}Z+(\bar Z
Z)\omega_{ij}\gamma^{ij}Z \ee
Applying the same identity  to the second term of this expression, we get  the final expression  for the transformation law of the
spinor $Z$
\be \delta Z=-(\bar
Z\gamma^{ij}Z)\omega_{ij}Z+(ZC\gamma^{ij}Z)\omega_{ij}C\bar Z,\quad
C=\left(\begin{array}{cccc}
 0 & 1 & 0 & 0\\
 -1 & 0 & 0 & 0\\
 0 & 0 & 0& 1\\
 0 & 0 & -1& 0
    \end{array}
\right) \label{finald6}
\ee
Here we used identities
 \be (\bar Z\gamma^l
Z)\omega_{ij}\gamma^{ij}\gamma^lZ=(\bar Z
Z)\omega_{ij}\gamma^{ij}Z,\quad \frac{1}{4}(\bar
Z\gamma^{lm}Z)\omega_{ij}\gamma^{ij}\gamma^{lm}Z=-(\bar Z
Z)\omega_{ij}\gamma^{ij}Z. \ee
To make the origin  of transformation (\ref{finald6}) more transparent, let us reformulate  it in terms of quaternions.
Taking into account the definition of spinor $Z$ and the block-diagonal form of matrix $C$ we introduce the notation

\be
{\bf u}_1=z_1+ z_2{\bf j}, \quad {\bf u}_2=z_3+ z_4{\bf j},\quad \btau=(\bar
Z\gamma^{ij}Z)\omega_{ij}+ (ZC\gamma^{ij}Z)\omega_{ij}{\bf j},
\ee
In these terms the transformation (\ref{finald6}) reads
\be
\delta {\bf u}_\alpha={\btau}{\bf u}_\alpha\;.
\label{q6}\ee
Since $\btau$ is pure imaginary quaternionic function, we conclude, that finite transformation corresponding to (\ref{q6}) is of the form
\be
{\bf u}_\alpha\to{\bf G}({\bf u},{\bf \bar u}){\bf u}_\alpha,\qquad {\bf G}{\bf \bar G}=1,
\label{q6g}\ee
i.e. we arrived  $SO(3)=S^3$ transformation given by (\ref{G}).

Similar to the $d=3+1$ case, with an appropriate redefinition of the parameters $\omega_{ij}$, the transformation (\ref{q6})
(and (\ref{q6g})) can be ``flattened''.

The metrics, which are invariant under such a ``local" $SU(2)$ transformation will be defined by the same formulae, as in the
$d=3+1$ case, (\ref{h}),(\ref{hol}),  where the complex spinor $Z^\alpha$ is replaced by the quaternionic one ${\bf u}^\alpha$.

The reductions of the Lagrangians  constructed by the use of this $SU(2)$-invariant metric could be done
following the receipt presented in \cite{gknty} for the reduction  by the action of the ``flat" $SU(2)$ transformation.
In contrast with $d=3+1$ case, the reduced system will be depended not only on momenta $p_i$, but from the two
spin variables as well.

\section{$d=9+1 $ and third Hopf map}
Now, let us consider the action of the little group of momentum for the last, $d=9+1$ dimensional case, corresponding to the third Hopf map.

For the $SO(1,9)$ algebra the minimal spinor is a Majorana-Weyl one.
Choosing the chiral representation of $\Gamma$ -matrices
\be \Gamma^0= \left(
\begin{array}{ccc}
 0 & \imath{\bf 1}_4\\
-\imath{\bf 1}_4 & 0
\end{array}
\right),\quad \Gamma^i=\left(
\begin{array}{ccc}
 0 &  \gamma^i\\
\gamma^i & 0
\end{array}
\right) \ee one can write the 32-spinor $\lambda$ in the following form
\be
\lambda=(Z_1,C_4\bar Z_1,Z_2,C_4\bar Z_2,0,....,0),\qquad {\rm
where}\quad
 Z_1=(z_1,z_2,z_3,z_4),\quad Z_2=(z_5,z_6,z_7,z_8)\ee
are general 4-dimensional Dirac spinors and $C_4$ is   $(4\times 4)$ C-matrix for dimensionality $4+1$.

Similar to  the previous sections, we define the quantities $p_\mu$ by the expression (\ref{pdef4}).
Denoting  the first sixteen elements
of spinor $\lambda$ by $Z$, we will get,  for  the quantities $p_\mu$, the realization  (\ref{pthz}).

Then, performing manipulations, which are completely similar to those in  the Fourth and Fifth  Sections,
we shall get the following expression for the transformations,  which leave invariant the momenta  $p_\mu$:
\be \delta
Z=-2\frac{\omega_{ij}(\bar Z\gamma^j
Z)}{p_0}\gamma^{i}Z+\omega_{ij}\gamma^{ij}Z \label{initd66}\ee
This expression can be transformed using Fierz reordering formula.
Namely, for any $(16\times16)$ matrix  $A$ we have following decomposition
\be A=\frac{1}{16}Tr(A){\bf
1}_{16}+\frac{1}{16}Tr(A\gamma^i)\gamma^i-\frac{1}{32}Tr(A\gamma^{ij})\gamma^{ij}-\frac{1}{96}Tr(A\gamma^{ijk})\gamma^{ijk}+\frac{1}{384}Tr(A\gamma^{ijkl})\gamma^{ijkl},
\ee where \be
\gamma^{i_1i_2,...i_k}=\frac{1}{k!}\gamma^{[i_1}\gamma^{i_2}...\gamma^{i_k]}
\ee
Applying twice this identity to (\ref{initd66}),  we get the
 simple final expression \be
 \delta Z=-\frac{1}{6}\omega_{ij}\left(ZC\gamma^{ijlm}Z\right)\gamma^{lm}Z
\label{octfin}\ee
Taking into account general considerations of the Second Section, we conclude, that (\ref{octfin}) defines the $S^7$
action in the fiber of third Hopf fibration $S^{15}/S^{7} =S^8$. Thus, being reformulated in octonionic terms,
it could be represented
in the form
\be {\bf u}_\alpha \mapsto ({\btau}{\bf g})({\bf
\bar g}{\bf u}_\alpha)= {({\btau}{\bf u}_1)({\bf
\bar u}_1{\bf u}_\alpha)} ,\qquad {\rm Re}\;\btau\;=0, \label{octtranslocal} \ee
which is an infinitizimal form of (\ref{octtrans}).
%
In contrast with (\ref{1f4}) and (\ref{finald6}), transformation (\ref{octfin}) does not possess Lie-algebraic structure,
but form some quadratic algebra. This is in accordance with the fact, that $S^7$ is not a group manifold.
It is clear from the expression (\ref{octtranslocal}), that this transformation can not be flattened in analogy with (\ref{flat4}).
Consequently, it is not so clear for us, how to construct the octonionic ($S^7$-invariant) analogs of the
one-form (\ref{1formGlob}) and metric (\ref{metricGlob}) (which are invariant under ``flat" $S^1$/ $S^3$ transformations).
The extension of the reduction procedure developed in \cite{gknty}, to the given octonionic case is also not so straightforward.
In any case, consideration of these problems should definitely be easier to consider in the complex terms (\ref{octfin}),
than in octonionic ones.
 We suppose to consider them  elsewhere.

\section{Conclusion and Outlook}

We show how Hopf maps are connected to the little groups
 of ``preon'' orbits in relevant dimensions $1+3$, $1+5$ and $1+9$.
  Simple realization of Hopf maps for first two cases, known earlier \cite{lnp},
  is recovered from the present approach. The unsolved challenge remains construction
  of (necessarily non-quadratic) Lagrangian, suited for the reduction, realizing third Hopf map,
  which will bring to the still unknown mechanical system, generalizing those with monopoles for lower dimensions.
  For that case we get a simple expressions for the action of little group on a $S^7$ fiber, which presumably will help
  in construction of  Lagrangian, necessarily non-quadratic, realizing the third Hopf map.  Another direction of study may be
  to generalize Poincar{\'e}, i.e. to substitute  Lorentz group, as well as tensorial central charges, with some more general objects,
  containing those as a subgroups, with purpose to obtain generalization of little groups and corresponding Hopf maps. Very deep such a
  generalization is known  already few years, that is a Kac-Moody
  approach to superstrings/supergravity theories (see \cite{west}, \cite{nicolai}). For our purposes we need an analog of our spinor $\lambda$ in that approach. Its explicit description appears to be one of the main problems of that theories, such a ``spinor representation'' would be a supersymmetry charge, important for construction of full supersymmetric theory. First few therms of decomposition of that charge w.r.t. the Lorentz group are known, only. \\

{\large Acknowledgments.}
 We are  grateful to Tigran Hakobyan   for useful comments. The work was supported by
 and ANSEF-2229PS grant and by
 Volkswagen Foundation  grant I/84~496.

\end{document}